\documentclass[onecolumn]{aa}
\usepackage{graphicx}
\usepackage{txfonts}

\begin{document}

\title{On the excess of power in high resolution CMB experiments}
\author{J.M. Diego\inst{1} 
        P. Vielva\inst{2} 
        E. Mart\'\i nez-Gonz\'alez\inst{3} 
        J. Silk\inst{4}
        }

   \offprints{J.M. Diego}

    \institute{University of Pennsylvania, Philadelphia, PA 19104, USA.\\
               \email{jdiego@physics.upenn.edu}
               \and
                Physique Corpusculaire et Cosmologie, Coll\`ege de France, 
                11 pl. M. Berthelot, F-75231 Paris Cedex 5, France.
               \and
                Instituto de F\'\i sica de Cantabria, Avda. Los Castros 39005,Santander, Spain.
               \and
                University of Oxford, Astrophysics Department, 1 Keble Road, Oxford OX1 3RH, UK.
              }

\date{Draft version \today}

\abstract{
  We revisit the possibility that an excess in the CMB power spectrum 
  at small angular scales (CBI, ACBAR) can be due to galaxy clusters 
  (or compact sources in general). 
  We perform a Gaussian analysis of ACBAR-like simulated data 
  based on wavelets. 
  We show how models with a significant excess should show a clear  
  non-Gaussian signal in the wavelet space. In particular, a value of the 
  normalization $\sigma_8 = 1$ would imply a highly significant skewness and 
  kurtosis in the wavelet coefficients at scales around 3 arcmin. Models 
  with a more moderate excess also show a non-Gaussian signal in the simulated data. 
  We conclude that current data (ACBAR) should show this signature if the excess is 
  to be due to the SZ effect. Otherwise, the reason for that excess should be 
  explained by some systematic effect.  
  The significance of the non-Gaussian signal depends on the cluster 
  model but it grows with the surveyed area. Non-Gaussianity test performed 
  on incoming data sets should reveal the presence of a cluster population even for 
  models with moderate-low $\sigma_8$'s.


  }

\maketitle


\section{Introduction}\label{section_introduction}
In the last couple of years there have been several results suggesting that 
there is a significant excess in the power spectrum of CMB 
experiments at small scales ($\theta < 5$ arcmin or $\ell > 2000$) 
(CBI, Pearson et al. 2003; ACBAR, Kuo et al. 2002). 
The best candidate for this excess (if confirmed) is the SZ effect 
signal in galaxy clusters. 
We expect the excess in power due to galaxy clusters to be observed 
at some point at scales smaller than $\approx$ few arcmin 
(or $\ell > 3000$) although the exact value is model dependent. 
However, the amplitude of the excess claimed by recent experiments 
is larger than what it is expected for the current most fashionable 
models. In fact, if that excess is confirmed to be due to galaxy clusters, 
it would require high values for the normalization of the matter power 
spectrum ($\sigma_8 \approx 1$) (e.g Bond et al. 2002). 
These high values of $\sigma _8$ would contradict the values 
derived directly from observations of galaxy clusters 
(e.g Efstathiou et al 2002).
Another contribution to the excess in power could be due to unresolved 
non-subtracted point sources.
It is expected to be more important for experiments
at low frequencies in the microwave band (like CBI) than at frequencies 
around 150 GHz (like ACBAR). Since a strong contribution from radio 
sources is expected in the former case and in the later 
the extragalactic point source contribution is at the lowest level along the 
microwave band, a more significant residual after the subtraction/estimation 
analysis can be present at centimeter rather than millimeter wavelengths. 
Moreover, the effect of clustering of point sources should be considered 
in the estimation of the residual (Toffolatti et al. in preparation).\\
An interesting aspect of the excess in power is that no compact sources 
(clusters or point sources) are clearly seen directly in the data, 
thus suggesting that the origin  of this excess may be due to a different 
nature other than the compact sources. 
One has only to realize that the power 
spectrum is an averaged quantity over the surveyed area and that 
the compact (bright) sources are not distributed over the whole 
area.
As a consequence, many pixels will enter in the average with 
negligible values.
On the other hand,  the CMB does distribute over the whole surveyed area. 
This implies that if the power spectrum of compact sources dominates the 
power spectrum of the CMB at small scales, then those sources should be 
seen in the map at small scales. Then, where are they ? 
The debate is still open and it is not clear whether the excess is 
due to compact sources or is just a systematic effect.\\
In this paper we will study the SZ contribution to the excess found 
in the power spectrum and its cosmological implications. In particular we will 
analyze the Gaussian deviations introduced in the wavelet coefficients. 
Since the CMB has not shown (up to date) any departures from Gaussianity at 
least at those small scales (for larger scales see Vielva et al. 2003) 
therefore any non-Gaussian detection would point out the presence of compact 
sources. 

The analysis presented in this paper will be useful to confirm or 
reject the hypothesis that the excess is due to galaxy clusters in 
current CMB data. The analysis would be also a desirable consistency 
check with the near future incoming data sets. \\
The reader will find very interesting for instance the pioneering work of 
Aghanim \& Forni (1999) and the more recent of Rubi\~{n}o-Martin \& 
Sunyaev (2003) where similar questions are considered.

\section{Searching for compact sources signatures}
\begin{figure*}
   \centering
   \includegraphics[width=17cm]{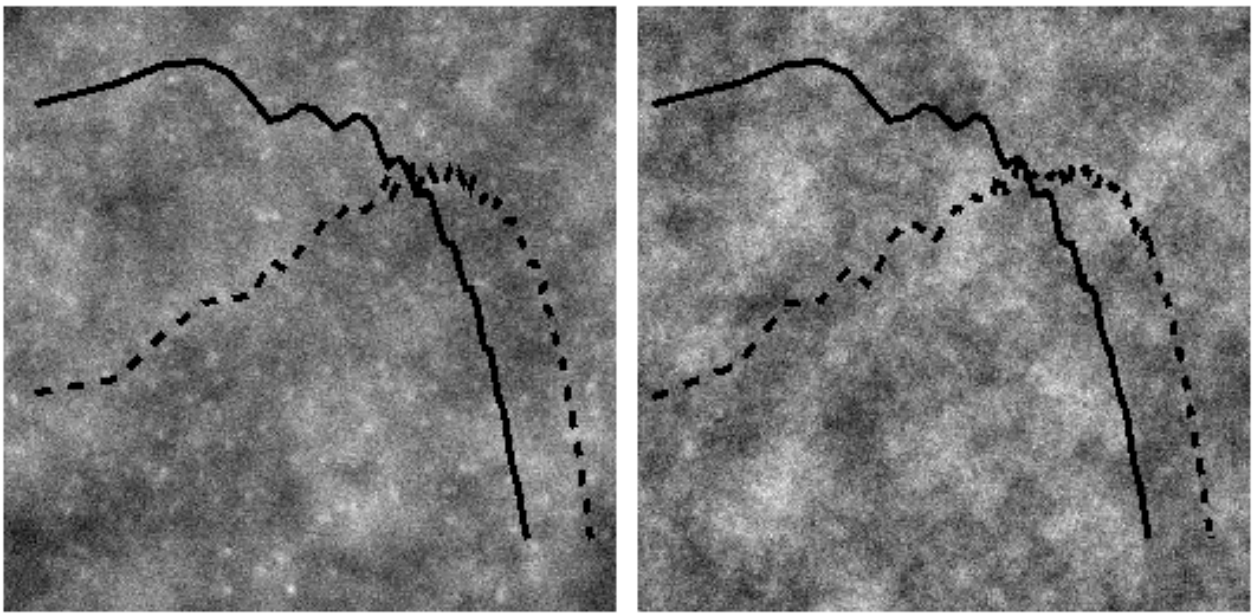}
   \caption{
            The power spectrum of the CMB realization and 
            the point sources (convolved) is shown as solid and 
            dashed lines respectively.  
            The population of sources consist of 1000 point sources 
            distributed in 1.5 \% of the total number of pixels and 
            with random (uniform) amplitudes  
            between 0 and $1 \times 10^{-3}$K (left) 
            and 30000 point sources (in 40 \% 
            of the total number of pixels) with random 
            amplitudes (uniform) between 0 and $2.5 \times 10^{-4}$ K (right). 
            The point where 
            both powers cross is at about $\ell \approx 2000$. 
            Each image is $6 \times 6$ sq. deg.
           }
   \label{fig_0a}
\end{figure*}

Our aim is to show how it should be possible to reject the 
possibility that the excess is due to galaxy clusters or to 
confirm that the excess in power is due to a signal which is 
compact. The main idea follows the argument in the introduction,
{\it if the power spectrum at small scales is dominated by compact 
(non-Gaussian) sources, then these compact sources should produce a 
non-Gaussian signal at small scales}. \\
We illustrate this point with figure \ref{fig_0a} (left) where we simulate 
a realization of the CMB (with a standard WMAP-like power spectrum,
Hinshaw et al. 2003) and we add a population of point sources 
(we also add instrumental noise with an rms of $30 \ \mu$K per pixel). 
The population of sources consist of 1000 point sources (convolved with 
an antenna beam of 5 arcmin) with amplitudes ranging from 0 to 0.001 K (uniformly 
distributed in that range). In the same plot we also show the power 
spectrum of both the antenna-convolved CMB and the convolved point sources. 
At $\ell \approx 2000$ the power spectrum of the point sources starts 
to dominate the power spectrum of the CMB. Many point sources can also 
be seen very clearly in the map even without any filtering. 
There is however other situation in which an excess in the power 
due to compact sources does not mean one actually has to see the sources 
in the map clearly. An excess in power can be due to the presence 
of relatively few bright sources (this is the case of figure 
\ref{fig_0a}) or it can be due to the presence of a large number 
of weak, almost undetectable compact sources. The later case 
is shown in figure \ref{fig_0a} (right). In this case we take 30000 point 
sources (instead of the 1000 point sources of figure \ref{fig_0a} left) 
but we lower the maximum amplitude down to 
$2.5 \times 10^{-4}$K. This second case could explain why 
the experiments claiming the excess in power have not found 
the sources of this excess. 
However, in both cases, if the excess is due to compact sources, 
their signal must be detectable not only in the frequency domain 
(power spectrum) but also in the real space domain (at the scales where they 
dominate the power). 
A good way of looking for signatures of the compact sources in real space 
is by looking at the Gaussian properties of the data. Since compact sources 
are known to be very non-Gaussian signals, they should leave a 
non-Gaussian imprint in the data. 
The problem is that if one looks directly at the Gaussian properties 
of the data, the statistics may be easily dominated by the properties 
(Gaussian) of the CMB and the noise. This will be particularly true in 
our case where we may expect a small fraction of the total number of pixels 
to be affected by the non-Gaussian compact sources.  
A better approach is to pre-process the data somehow to increase the 
signal-to-noise ratio of the non-Gaussian compact sources with respect 
to the Gaussian CMB and noise.
This later point can be achieved by doing an appropriate filtering of 
the maps. 
When we look at the Gaussian features of our toy model in figure \ref{fig_0a}, 
we find that after a proper filtering they show non-Gaussian signatures which correspond to the 
presence of the compact sources. For instance, looking at the skewness with the MHW (Mexican Hat Wavelet) 
at scales of 1.5 pixels there is a non-Gaussian signal at 40$\sigma$ significance in the first case 
(fig. \ref{fig_0a} left) 
and at 5$\sigma$ significance in the second case (fig \ref{fig_0a} right). 
The significance is calculated from a set of 500 simulations 
with the same CMB power spectrum and noise as in the cases of figure \ref{fig_0a}. \\

There are a variety of filters which can make the job. In this work 
we will use the MHW (Cay\'on et al. 2000 and Vielva et al. 2001) although the 
reader is free to consider other (maybe more effective) filters 
(Sanz et al. 2001 and Herranz et al. 2002). \\
To quantify the contribution of the non-Gaussian compact sources 
we will use non-Gaussianity estimators like the skewness and kurtosis 
and we will apply these estimators to the wavelet-filtered data.  
We expect the signal from clusters to dominate the CMB signal 
only at small scales. 
Going to the wavelet space (or filtering with an appropriate filter) 
is important to select the scales of interest. 
Since the data is not available we will use simulations which reproduce 
the data to the best of our knowledge.\\ 

\section{Simulations}
We will try to emulate the ACBAR data, in particular we simulate the 
deepest section of the CMB5 field. It covers an area of the sky of about 
3 square degrees with an almost homogeneous noise level over 
the field of 8 $\mu$K per resolution element. 
We will include also the effect of the antenna beam by 
convolving our simulations with a 5 arcmin FWHM Gaussian beam.\\
The ingredients of the simulation will be the CMB component with 
a power spectrum which matches the WMAP power spectrum
(Hinshaw et al. 2003), the noise 
level assuming it is white-Gaussian and with the above level, and 
a population of compact sources (clusters) with a power spectrum which 
is consistent with the excess in power suggested by the ACBAR 
and CBI teams. \\

The galaxy cluster population is model dependent. We will assume 
two models which are marginally consistent 
with the measured power spectrum by ACBAR and CBI. 
In this paper we do not pretend to make a detailed study of the models 
which can explain the excess. We will show however how models predicting 
an excess in the range of the observed one should also produce non-Gaussian 
signals at small scales. A more detailed modeling would be required however, if a 
serious analysis of the real data is performed. Unfortunately, the data is not 
freely available. Nevertheless, our two simple models should suffice to describe 
the range of possible models explaining the excess. 
 
\begin{figure}
   \centering
   \includegraphics[width=12cm]{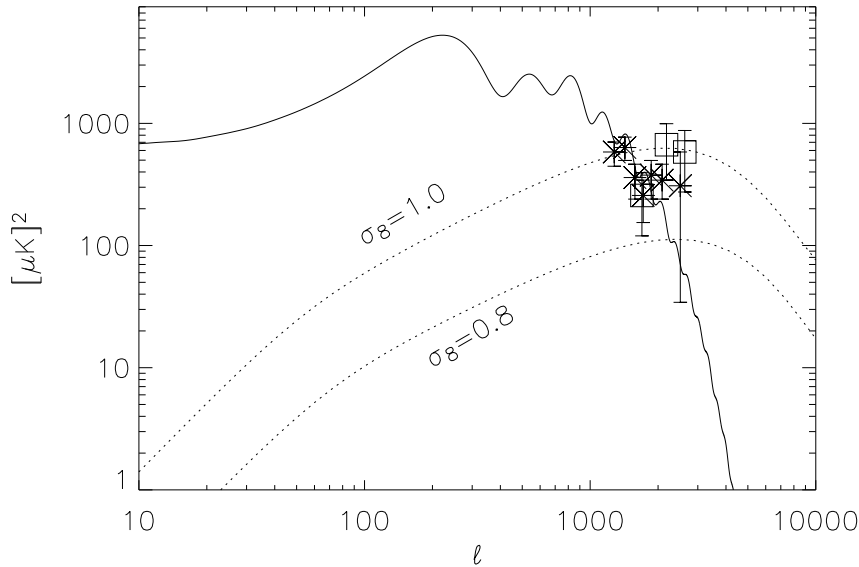}
   \caption{
            Power spectrum of the CMB used to simulate the data 
            (solid line) compared with the power spectrum of the 
            two models used to simulate the clusters (dotted lines). 
            Also shown is the ACBAR (stars) and CBI (squares) data showing 
            the excess in power at high $\ell$'s 
           }
   \label{fig_4}
\end{figure}

We will call model A (or $\sigma _8 = 1.0$) the one with a significant  
excess and marginally 
consistent with the upper limit ($\approx 2\sigma$) of the 
data error bars. Model B (or $\sigma _8 = 0.8$) is marginally 
consistent ($\approx 2\sigma$) with the lower limit of the error bars. 
For modeling the clusters we will use Press \& Schechter (1974) 
and a parametrization of the temperature-mass and the virial 
radius-mass relations following Diego et al. (2001). 
We also assume a $\beta$-model for the electron density ($\beta = 2/3$). 
The details of the model are not particularly relevant as long as their power 
spectrum is consistent with the excess. 
The parameters of the models are described in table 1. 
There is an extra-parameter not listed in the table which 
controls the ratio $p=R_v/R_c$ which we fix to $p=10$ ($R_v$ and $R_c$ 
are the virial and core radius respectively). In both cases, we assume 
a $\Lambda$CDM flat universe with $\Lambda = 0.7$.
The power spectrum of the two models is shown in 
figure \ref{fig_4}. 
The difference in the cosmological parameters between the two models 
is just the normalization parameter $\sigma _8$. We also have changed 
the temperature normalization which boost the power in model A by a 
factor $(9/8)^2$ with respect to model B. Changing the rest of the 
parameters can also change the power (specially at small scales). 
For our purposes, models A and B are sufficient since they represent 
the upper and lower limits respectively given by the data. 

\section{Results}
\begin{figure}
   \centering
   \includegraphics[width=12cm]{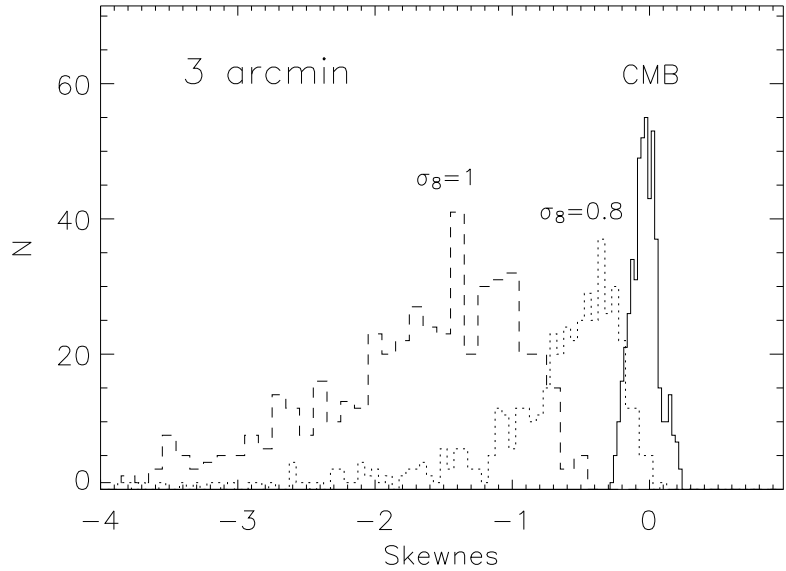}
   \caption{
            Distribution of skewness of the simulated maps after filtering 
            with a MHW of 3 arcmin scale.
           }
   \label{fig_1}
\end{figure}

\begin{figure}
   \centering
   \includegraphics[width=12cm]{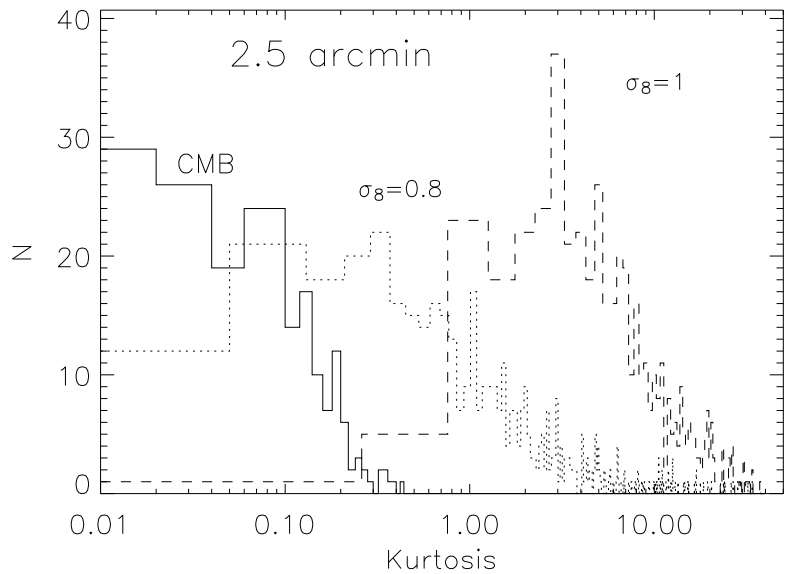}
   \caption{
            Distribution of kurtosis of the filtered maps after 
            filtering with a MHW of 2.5 arcmin scale.
           }
   \label{fig_2}
\end{figure}

With the above models, we make 500 simulations of the CMB plus noise 
and 500 simulations for each of the two models (Model A with 
$\sigma _8 = 1$, and model B with $\sigma _8 = 0.8$). 
With these simulations we build 3 kinds of maps, 
i) CMB + Noise, ii) CMB + Noise + Model A, and iii) CMB + Noise + 
Model B. 
The first set of maps is necessary to establish the significance of 
any possible non-Gaussian detection. With the second and third sets of maps 
we will test the sensitivity of the skewness and kurtosis of the 
filtered maps to the non-Gaussian signatures due to the clusters.
We use the MHW to select different scales in the simulated maps
and to amplify the non-Gaussian signals with respect to the Gaussian 
ones (CMB and noise). Then we look at the skewness and kurtosis of 
the wavelet-filtered maps as a 
function of the scale. Eventually, for a range of scales we can see 
the non-Gaussian features of models A and B and there will be an optimal 
scale for which these features are seen most clearly. 
In figures \ref{fig_1} and \ref{fig_2} we show the skewness and 
kurtosis for the two corresponding optimal scales. 
The most remarkable thing from these results is that for models 
having a high power at small scales (like Model A), we should expect 
to see a significant non-Gaussian signature in maps like the ACBAR 
ones. \\
\begin{table} 
 \label{table1}
 \begin{center}
  \begin{tabular}{|c|c|c|c|c|c|c|c|}
    \hline
     &$\Omega _m$&$\sigma _8$&$T_o$&$\alpha$&$\phi$&$R_o$&$\psi$ \\
    \hline
    \hline
     A&0.3&1.0&9.0&0.55&1.0&1.5&-1\\
    \hline 
     B&0.3&0.8&8.0&0.55&1.0&1.5&-1\\
    \hline 
  \end{tabular}
  \caption{Parameters taken in models A and B. The temperature-mass 
           relation is given by $T = T_o M_{15}^{\alpha} (1+z)^{\phi}$ 
           and the virial radius-mass relation is given by 
           $R_v =R_o M_{15}^{1/3} (1+z)^{\psi}$.   
           All numbers are dimensionless except $T_o$ (Kev) and 
           $R_o$ ($h^{-1}$ Mpc)}. 
 \end{center}
\end{table}
Even for models with a relatively low power (model B) we could expect 
to see non-Gaussian features in an appreciable fraction of the 
realizations. \\
At this point, it is important to note that the dispersion of 
the distributions shown in figures \ref{fig_1} and \ref{fig_2} 
decreases as the surveyed area increases. This means, that a larger 
area of the sky would reduce the overlapping area between the  
distributions of model B and the Gaussian case, thus allowing a detection of 
the non-Gaussianity induced by clusters even for these models.

\section{Discussion}
We have presented a small (but interesting) excercise showing how current data 
(ACBAR CMB5 field) is expected to have non-Gaussian signals if the claimed excess 
in the power spectrum is fully due to the SZ effect. With this excercise we want to 
send a brief but clear note to the community. Current data should be checked for 
non-Gaussian signals at small scales before any excess in the power spectrum is 
claimed as due to the SZ effect.

The potentiality of this approach is based on 
the fact that, up to date, no intrinsic deviations from Gaussianity have 
been found in the CMB at these small scales. Any deviations would 
indicate the presence of compact sources which are
expected to be the dominant foreground at arcmin scales.\\
We have presented the result of the level of non-Gaussianity we 
should expect in ACBAR-like data if a significant excess in power 
is due to galaxy clusters. 
Given the uncertainties in the estimation of the power 
spectrum by ACBAR and CBI, we explore two models which are in the 
lower and upper limit of these uncertainties and we find that for the 
model in the upper limit we should expect a clear non-Gaussian signal 
at MHW scales around 3 arcmin (the Gaussian and A-model distributions 
in figures \ref{fig_1} and \ref{fig_2} overlap by less than 1 \%). 
In the other case (lower limit B-model), we should expect a non-Gaussian 
signal in $\approx 75 $\% of the cases if we look at the kurtosis and 
in $\approx 80 $\% if we look at the skewness (both at $> 99$ \% 
confidence limit). 
By performing this analysis on the ACBAR data (for example), one 
could confirm or reject several galaxy cluster models.  \\
Although we have based our simulations on galaxy clusters, a similar 
argument could be made if the excess is due to point 
sources. A way to distinguish between both would be the sign of 
the skewness of the MHW coefficients: a positive sign would point 
out to point sources whereas a negative one would imply the presence 
of galaxy clusters (see Rubi\~{n}o-Martin \& Sunyaev 2003).
For experiments where the level of non-Gaussianity is expected 
to be small, a more efficient manner to discriminate between the CMB 
and galaxy clusters is by defining a Fisher discriminant which contains 
information about both the skewness and kurtosis and also at several scales 
(Mart\'\i nez-Gonz\'alez et al. 2002). 
These techniques will prove to be very useful with the near future data.

We would like to conclude by being provocative by insisting that actual data 
should be checked for non-Gaussian signals before suggesting that the excess is due  
to the SZ effect. If the data is found to be compatible with Gaussianity, then the 
systematics may be the reason for that excess. For instance, the power spectrum 
may have been overestimated at small scales if the covariance matrix of the instrumental noise 
is slightly undetermined at those small scales. Although we are not the first suggesting 
that non-Gaussianity studies are useful to detect SZ signatures (See Aghanim \& Forni 1999 
or  Rubi\~{n}o-Martin \& Sunyaev 2003), we are the first in predicting 
that if a significant excess claimed by recent experiments is due to the SZ effect, 
in general, non-Gaussian signals should be found with a high significance 
in at least one of those data sets, the CMB5 field of ACBAR. 
The reader may consider our approach too simplistic but we think is important to 
highlight our main point with simple arguments, current data should be checked for 
non-Gaussian signals.

\section{Acknowledgments}
We would like to thank S. Majumdar, Max Tegmark 
L. Toffolatti and J. Gonz\'alez-Nuevo for useful comments.
This work was supported by the David and Lucile
Packard Foundation and the Cottrell Foundation. 
We thank the RTN of the EU project HPRN-CT-2000-00124. PV acknowledges support from 
IN2P3 (CNRS) for a post-doc fellowship and EMG acknowledges
partial support from the Spanish MCYT project ESP2002-04141-C03-01. 




\begin{thebibliography}{2001}
%
\bibitem{Aghanim1999} Aghanim N, Forni O., 1999, A\&AS, 347, 409.

\bibitem{Bond2002} Bond et al. 2002. Preprint astro-ph/0205386.
%
\bibitem{} Cay{\'o}n L., Sanz J. L., Barreiro R. B., Mart{\'\i}nez--Gonz{\'a}lez E., 
Vielva P., Toffolatti L., Silk J., Diego J. M. \&  Arg{\"u}eso F., 2000, MNRAS, 315, 757
%
\bibitem{b7} {Diego J. M., Mart{\'\i}nez-Gonz{\'a}lez E., Sanz J. L.,
Cay{\'o}n L. \& Silk J., 2001, MNRAS, 325, 1533}
%
\bibitem{Herranz} {Herranz D., Sanz J. L., Barreiro R. B. \&
Mart{\'\i}nez-Gonz{\'a}lez E., 2002b, ApJ, 580, 610}
%
\bibitem{efstathiou2002} Efstathiou G., 2002, MNRAS 330, L29.
%
\bibitem{} {Hinshaw G. et.al., 2003, ApJS, 148, 135}
%
\bibitem{acbar} {Kuo C. L. et al., 2002, ApJ submitted, preprint astro-ph/0212289}
%
\bibitem{SMHW-SHW:PLANCK} Mart{\'\i}nez--Gonz{\'a}lez E. Gallegos J. E. Arg{\"u}eso F. 
Cay{\'o}n L. \& Sanz J. L., 2002, MNRAS, 336, 22
%
\bibitem{CBI} {Pearson T. J. et al., 2003, ApJ, 591, 556}
%
\bibitem{Press74} Press W. H., Schechter P., 1974, ApJ, 187, 425.
%
\bibitem{Rubino2003} Rubi\~{n}o-Martin J.A., Sunyaev R.A., 2003, MNRAS. 344, 1155.
%
\bibitem{b21} {Sanz J. L., Herranz D. \& Mart{\'\i}nez-Gonz{\'a}lez E.,
2001, ApJ, 552, 484}
%
\bibitem{MHW} Vielva P., Mart{\'\i}nez--Gonz{\'a}lez E., Cay{\'o}n L., Diego J. M., 
Sanz J. L. \& Toffolatti L., 2001, MNRAS, 326, 181
%
\bibitem{NG} Vielva P., Mart{\'\i}nez--Gonz{\'a}lez E., Barreiro R. B., Sanz J. L.
\& Cay{\'o}n L., 2003, ApJ submitted, preprint, astro-ph/0310273
%
\bibitem{Zhang2003} Zhang Y-Y., Wu X-P., 2003, ApJ, 583, 529. 
%

\end{thebibliography}
\end{document}